\documentclass[preprint,12pt]{elsarticle}

\usepackage{amsmath,amssymb}
\usepackage{lineno}
\usepackage{gensymb}
\modulolinenumbers[5]

\journal{Journal of Magnetism and Magnetic Materials}

\begin{document}

\begin{frontmatter}

\title{Magnetic properties of wurtzite  {(Ga,Mn)As} }

\author[1]{Katarzyna Gas}

  \address[1]{Institute of Physics, Polish Academy of Sciences, Aleja Lotnikow 32/46, PL-02668 Warsaw, Poland.}
\address[2]{Department of Physics and Electrical Engineering, Linnaeus University, SE-391 82, Kalmar, Sweden}

\author[1,2]{Janusz Sadowski}

\author[1]{Maciej Sawicki\corref{mycorrespondingauthor)}}
\cortext[mycorrespondingauthor]{Corresponding author. Maciej Sawicki}
\ead{mikes@ifpan.edu.pl}

\begin{abstract}
Here we report on detailed  studies of the magnetic properties of the wurtzite (Ga,Mn)As cylindrical shells.
Ga$_{0.94}$Mn$_{0.06}$As shells have been grown by molecular beam epitaxy at low temperature as a part of multishell cylinders overgrown on wurtzite (Ga,In)As nanowires cores, synthesized on GaAs (111)B substrates.
Our  studies clearly indicate the presence of a low temperature ferromagnetic coupling, which despite a reasonably high Mn contents of 6\% is limited only to below 30~K.
A set of dedicated measurements shows that despite a high structural quality of the material the magnetic order has a granular form, which gives rise to the dynamical slow-down characteristic to blocked superparamagnets.
The lack of the long range order has been assigned to a very low hole density, caused primarily by numerous compensation donors, arsenic antisites, formed in the material due to a specific geometry of the growth of the shells on the nanowire template.
The associated electrostatic disorder has formed a patchwork of spontaneously magnetized (macrospin) and nonmagnetic (paramagnetic) volumes in the material.
Using high field results it has been evaluated that the total volume taken by the macrospins constitute about 2/3 of the volume of the (Ga,Mn)As whereas in the remaining 1/3 only paramagnetic Mn ions reside.
By establishing the number of the uncoupled ions the two contributions were separated.
The Arrott plot method applied to the superparamagnetic part yielded the first experimental assessment of the magnitude  of the spin-spin coupling temperature within the macrospins in (Ga,Mn)As,  $T_{\mathrm{C}}=28$~K.
In a broader view our results constitute an important contribution to the still ongoing dispute on the true and the dominant form(s) of the magnetism in this model dilute ferromagnetic semiconductor.

\end{abstract}

\end{frontmatter}



\section{Introduction}

Even though (Ga,Mn)As dilute ferromagnetic semiconductor (DFS) has already been studied for over two decades \cite{Ohno:1998_S,Lee:2009_MT,Sawicki:2018_PRB}, there are still some unexplored issues concerning this canonical DFS materials.
Typically (Ga,Mn)As had been obtained in the form of thin films grown by molecular beam epitaxy (MBE), i.e, in 2-dimensional geometry of planar layers and most of the investigations were devoted to such samples.
Much less attention was paid to lower dimensional structures implementing (Ga,Mn)As, namely quasi 1-dimensional nanowire (NW), structures.
Concerning the growth there was a hope that one could straightforward  capitalize on the extremely broad and extensive activity in the field of self-assembled growth of NWs from many different semiconducting materials \cite{Guniat:2019_CR}, especially GaAs \cite{Plante:2008_JCG} and other III-V semiconductors \cite{Barrigon:2019_CR}.
Those expectations proved futile, as specified below.

The growth of high structural quality GaAs NWs via typical Vapour-Liquid-Solid (VLS) mechanism requires rather high growth temperatures, in the range of $500 - 650$~\degree C \cite{Dubrovski:2006_PRE}.
Such conditions do not match those required for the MBE growth of (Ga,Mn)As, which has to be grown at much lower temperatures - typically in the range of $200 - 250$~\degree C for an Mn content above 1\%, the prerequisite condition for the ferromagnetic (FM) phase transition \cite{Ohno:1998_S}.
It became obvious that  the formation of (Ga,Mn)As in the 1-D geometry has to be obtained via other methods.
One of them relies on a "top-down" approach using nano-lithography and chemical etching to define stripes on planar (Ga,Mn)As layers.
This path proved successful in exploring magnetic and magneto-structural properties of (Ga,Mn)As at nano/micro-scale \cite{Ruster:2003_PRL,Figielski:2007_APL,Pappert:2007_NJP} and in elucidating some of its basic properties \cite{Neumaier:2007_PRL,Neumaier:2008_PRB}.
Yet for other investigations, where a high surface density of quasi 1-D nanostructures was essential, an elaboration of the bottom-up approach was the only possible option.
However, the MBE growth of (Ga,Mn)As NWs at low temperatures led to rather disordered, non-uniform NW assemblies \cite{Sadowski:2007_NL,Dluzewski:2009_JMicro}.
Another method - Mn ion implantation to GaAs NWs grown at optimum high temperature also resulted in  high density of structural defects and with no transition to FM phase \cite{Borschel:2011_NL}.
Finally, the MBE growth of Mn-doped GaAs NWs at higher temperatures, optimized for GaAs NWs resulted in the very low Mn content,  below $10^{18}$~cm$^{-3}$ \cite{Gas:2013_Nanoscale,Kasama:2015_JAP}.
Moreover, the accumulation of an excess Mn in the catalyzing droplets at the NW tops \cite{Kasama:2015_JAP} or as Mn-rich (MnAs) nanocrystals at the NW sidewalls was evidenced \cite{Bouravleuv:2013_JAP}.

Hence, the only way to obtain (Ga,Mn)As in the NW geometry using the bottom-up approach is to grow hybrid structures combining Mn-free NW templates grown at optimum conditions ($500 - 600$~\degree C in the case of GaAs), followed by low temperature growth of  (Ga,Mn)As over-shells at the primary NW sidewalls.
This approach was first successfully realized by Rudolph \emph{et al.}~\cite{Rudolph:2009_NL}, where about 30 nm thick (Ga,Mn)As shells were grown at low temperature on the {112} sidewalls of zinc-blende GaAs NWs previously grown at high temperature on GaAs(111)B substrate using Au-assisted VLS growth mode. The authors observed extremely narrow growth window ($\pm 10$~\degree C around the optimum temperature of about 234~\degree C) for uniform (Ga,Mn)As NW shell growth, and reported observation of very low temperature, $T < 20$~K, ferromagnetic-like  features.
Currently, the deposition of (Ga,Mn)As at the NWs sidewalls remains the only method able to produce high structural quality 1D structures of this DFS in large densities.
Simultaneously, a hexagonal wurtzite (wz) structure of the shell material can be forced, following the crystal structure of the NW template.

In order to obtain previously unknown (Ga,Mn)As in wz phase we have used purely wz III-V NWs as templates for shells deposition \cite{Siusys:2014_NL}.
Similarly to ref.~\cite{Rudolph:2009_NL} we also observed some FM-like features with a characteristic temperature of about 30~K.
These features were unambiguously assigned to a magnetic phase separation which took place in the  (Ga,Mn)As shells and the accompanying  dynamical slow down.
A clear formation of a spontaneous magnetization signalizing the occurrence of  the FM phase transition in thin wz (Ga,Mn)As shells has only been observed in the core-multishell NW heterostructures with additional (Ga,Al)As shell separating (Ga,Mn)As shells from (Ga,In)As NW cores \cite{Sadowski:2017_Nanoscale}.


\section{Samples and experimental}

\subsection{ Material preparation}

The core-multi-shell wz NWs investigated here are obtained from exactly the same process in which the strained (Ga,In)As-(Ga,Al)As-(Ga,Mn)As multi-shells were prepared to evidence the Curie temperature enhancement in MnAs NCs formed upon post-growth high temperature annealing \cite{Kaleta:2019_NL}.
The NWs are grown in the III-V MBE system (SVTA) using the VLS growth mode.
GaAs (111)B substrates piece with a thin layer of Au (5~\AA) deposited in another MBE system (dedicated for metals) was glued by In to molybdenum substrate holders.
Prior to the NWs growth the Au-covered GaAs substrate was heated to the temperature of the native oxide desorption (about 595~\degree C), monitored with reflection high energy electron diffraction (RHEED) system.
Next, the substrate temperature was raised by 10~\degree C and kept for 2 minutes.
At these initial preheating stages Au layer turns to Au-Ga eutectic nano-droplets  (liquid down to 350~\degree C \cite{Massalski:1987_B}).

The growth of NWs is initiated after lowering the substrate temperature to 550~\degree C.
At this moment a short-time growth of GaAs NWs is initiated at conditions optimized previously for wz NW phase i.e. with sufficiently high As/Ga flux ratio \cite{Dubrovskii:2017_CGD}.
Shortly after the appearance of  RHEED features typical for NWs (sharp spots slightly elongated in the direction perpendicular to the initial streaky patterns from the GaAs substrate), the GaAs growth is stopped and the substrate temperature lowered (in As flux) to about 500~\degree C for the subsequent (axial) growth of Ga$_{0.92}$In$_{0.08}$As NWs.
These core NWs are grown for 2 hours, what, at these growth conditions, results in about 2~$\mu$m long NWs, with surface density (defined by the initial Au droplets distribution) of about $2 \times 10^9$~cm$^{-3}$, as verified by scanning electron microscopy images of the final sample \cite{Kaleta:2019_NL}.
After completing the growth of (Ga,In)As NWs the substrate temperature is reduced to 450~\degree C and about 20~nm thick Ga$_{0.60}$Al$_{0.40}$As shells are deposited homogenously on all NW side-walls.
The uniform coverage is promoted by the substrate rotation during the shells growth.
Next, the substrate temperature is further reduced to about 200~\degree C, and nominaly 30~nm thick Ga$_{0.94}$Mn$_{0.06}$As shells are deposited.
Here the As flux is reduced to close-to-stoichiometric growth conditions, i.e.~to have  As/(Ga+Mn) flux ratio close to unity.
Finally, a thin (2~nm) low temperature GaAs capping shell is grown at the same conditions as those used for (Ga,Mn)As.
The final NWs have slightly tapered shapes with diameters changing from about 150~nm at the bottom (close to the substrate) parts to about 200~nm at the top parts \cite{Kaleta:2019_NL}.
The inverted tapering is due to shadowing effects during the shells deposition.

The elaborated multi-shell NW structure has been realized in order to achieve high structural quality needed to preserve the built-in strains required for Curie temperature engineering in MnAs nanocrystals \cite{Kaleta:2019_NL}.
The latter can be formed by post-growth high-temperature annealing (at about 450\degree C) \cite{De_Boeck:1996_APL,Sadowski:2017_Nanoscale}.
All the relevant structural characterization of the NWs studied here was given in \cite{Kaleta:2019_NL}.
In particular, the six-fold crystallographic symmetry of their crystal structure and the lack of any Mn-rich agglomerates in the as grown NWs was documented.

\subsection{ Sample preparation}

Despite the high crystalline quality and ordered form of the NWs, the as grown material is not suitable for direct magnetic measurements in integral magnetometers.
The reason is multi-fold.
The least problem constitutes the bulky GaAs substrate.
It certainly brings in the diamagnetic contribution whose magnitude dwarfs the signal of the about 20~nm thin (Ga,Mn)As shells.
At magnetic field $H = 50$~kOe and temperature $T = 5$~K the signal of the GaAs substrate exceeds that of the NWs by about  10 times.
It is generally taken for granted that this diamagnetic part can be subtracted in the whole $T$- and $H$-range by adequate scaling of the diamagnetism established at room temperature to match the specimen's weight and physical dimensions.
The latter takes into account size dependent strength of the sample-to-detection-coil coupling.
Some relevant guidance can be found in refs.~\cite{Stamenov:2006_RSI,Sawicki:2011_SST}.
This is however not the case of most of the semiconductor substrates since their native (lattice) diamagnetic susceptibility is slightly $T$-dependent and the total magnetic response depends on their purity.
The case of Al$_2$O$_3$ substrates has been recently discussed in ref.~\cite{Gas:2021_JALCOM}, GaAs in this respect has been examined in ref.~\cite{Ney:2006_JPCM}.
We note here that the problem of the detrimental contribution of $T$-dependent and non-linear in $H$ magnetic properties of common substrate materials can be substantially eliminated by the active, \emph{in situ} compensation, proposed recently by some of the present authors \cite{Gas:2019_MST}.

\begin{figure}[t!]
\begin{center}
\includegraphics[width = 8.0cm]{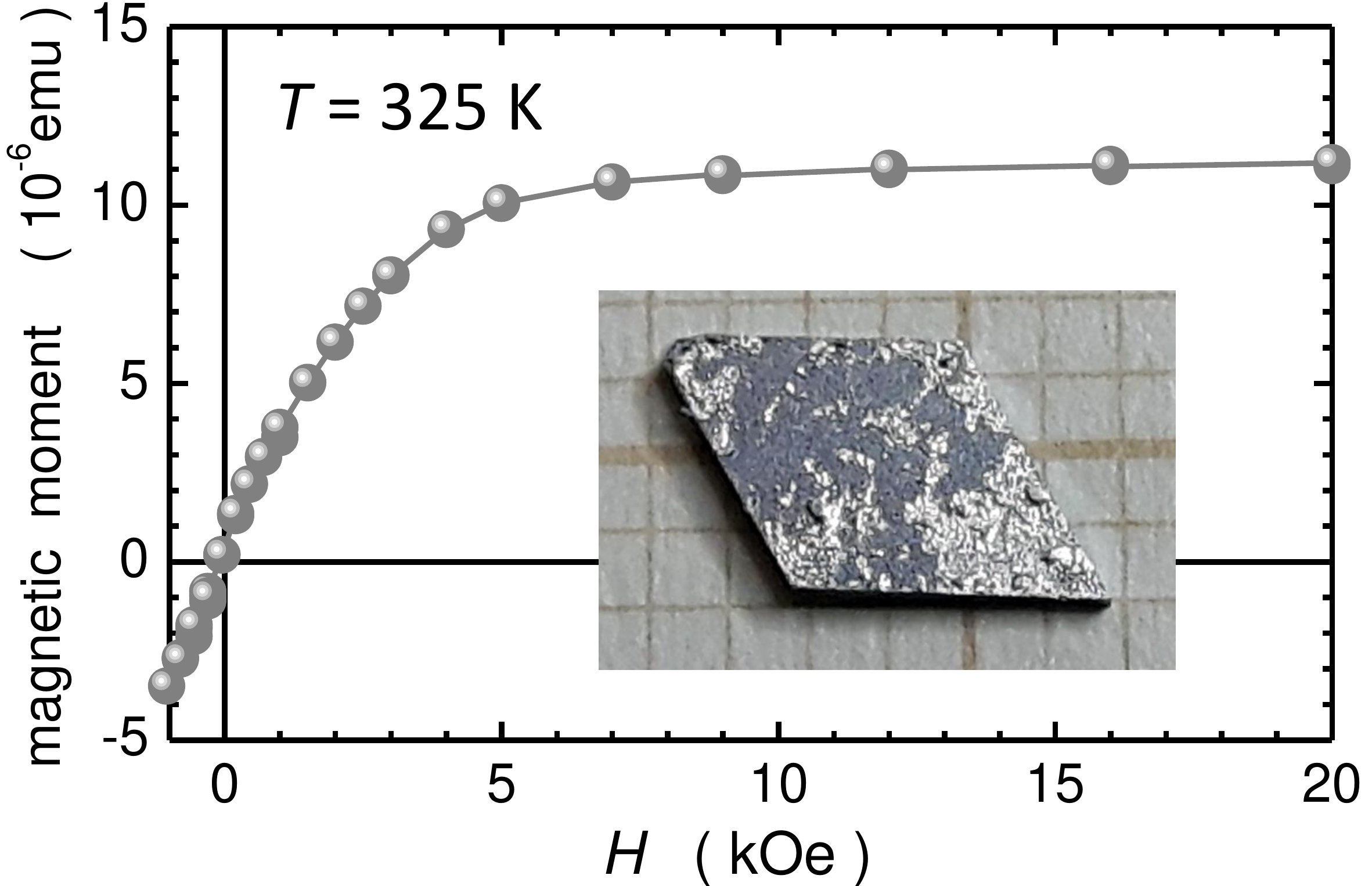}
\caption{(Color on line) Magnetic field, $H$, dependence of the signal brought about by ferromagnetic-like  contaminations present in the metallic glue (mostly In) used to fix and thermalize substrates in the MBE chamber. The back-side of the GaAs substrate investigated here is presented in the inset.  A linear in $H$, diamagnetic, contribution of the GaAs substrate has been subtracted.}
\label{Fig-In-glue}
\end{center}
\end{figure}

The second material issue is related to the substrate contamination occurring in the growth chamber by using various metallic "glues", aiming to both fix and thermalize the substrate to the substrate holder plate (molybdenum block) during the growth process.
Importantly, the problem does not stem from the use of In, Ga, their mixture, or other metal based substances - high purity components can be used.
The problem is that the fragments of these glues which are not obscured by the substrate absorb metal atoms used for the growth of the layer of interest, and so various ferromagnetic micro-particles can form, e.g.~Mn$_x$Ga$_y$, and many others.
They are then get transferred on the back-side of the substrate and its edges during the process of substrate detaching from the  molybdenum block  at a relevant temperature  (about 170~\degree C here).
We exemplify such a case in Fig.~\ref{Fig-In-glue}, where in the inset the typical distribution of the In-glue on the backside of the substrate is shown.
The main body of the figure presents the magnetic field dependence of the signal brought about by this metallic contamination (the diamagnetic contribution of the GaAs substrate has been subtracted).
It should be noted, by comparing the Y-scales on this and the figures presented below, that the magnitude of this spurious signal is comparable to the signals exerted by the NWs deposited on the front side of this specimen.
Most interestingly, the sigmoidal shape and a reasonable magnitude of this response could constitute a solid, however not-legitimate, basis to claim the existence of an above-room temperature ferromagnetism in the NWs. 
Usually, for the planar structures the metallic glue is removed by means of chemical etching or mechanical thinning using a corundum-like polishing powders.
The latter is most preferred since the In or Ga atoms are likely to diffuse deeply into the bulk of the substrate (even 200~$\mu$m is not uncommon).
These metallic residues are responsible for a disturbing superconducting contribution to the measured signal at relevantly low temperatures.
Neither of these methods is acceptable here, as both approaches will most certainly either disrupte the fragile NWs or simply destroy them all.

Importantly, the sample preparation method should also take care of the third detrimental issue.
It is the presence of (Ga,Mn)As-like material unintentionally grown in-between the NWs. 
Due to the shadowing effect during the NWs overgrowth it is expected that  neither the stoichiometry and the crystallographic structure nor the magnetic properties of this substance are known, so it cannot be eradicated from the magnetic data.
On the other hand both the thickness of this parasitic layer and its Mn contents should be similar to these wz-(Ga,Mn)As shells of interest, so this layers could exert a magnetic signal of a comparable magnitude to that of the NWs.

\begin{figure}[b!]
\begin{center}
\includegraphics[width = 8.0cm]{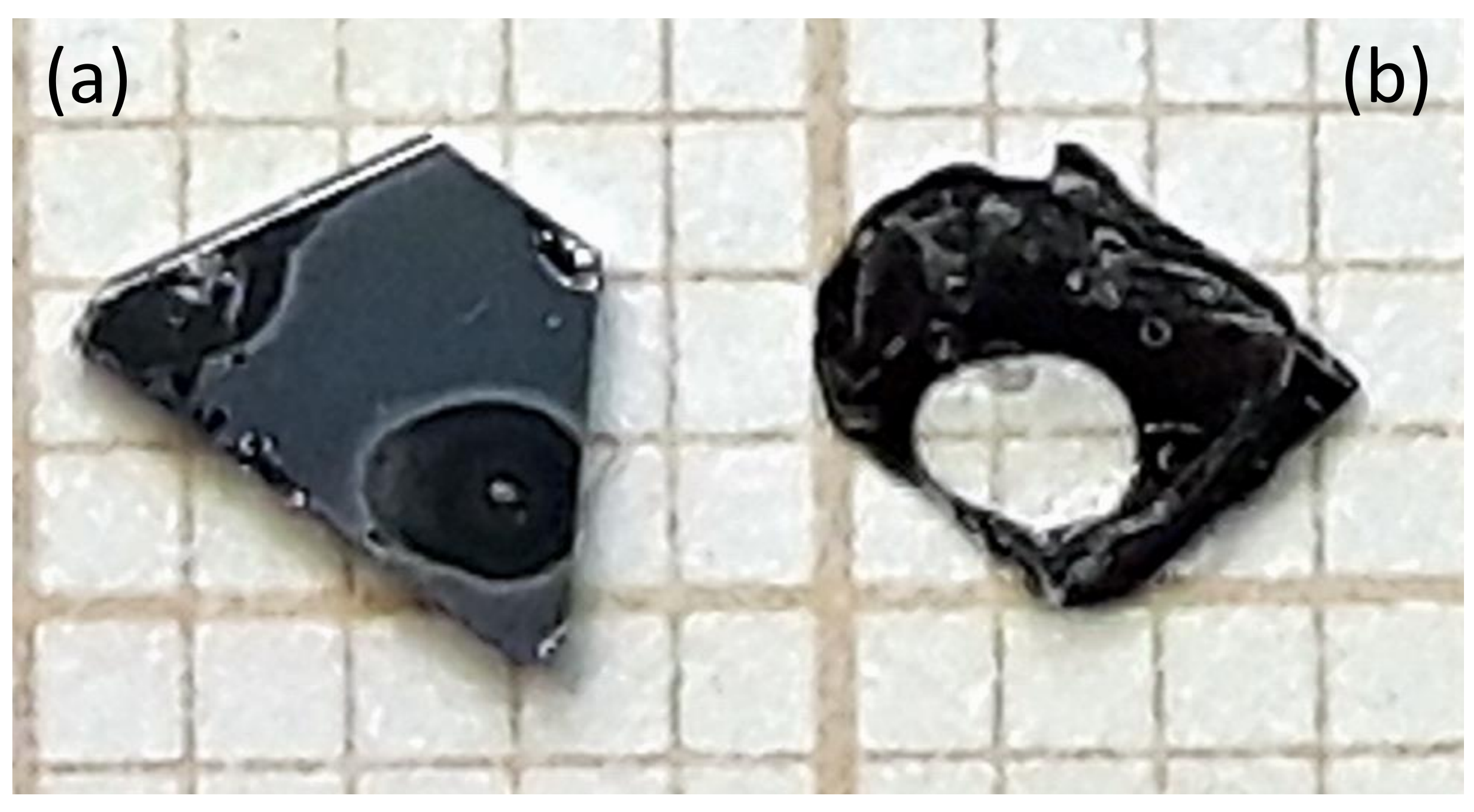}
\caption{(Color on line) The illustration of the method used to remove the nanowires (NWs) from the substrate. By a coverage by PMMA a rigid flake is formed (b), containing ordered NWs reaped at 77~K from the part of the substrate (a). The NWs constitute the charcoal-black substance under the transparent flake of PMMA. The specimen seen in (b) is investigated in this report. }
\label{Fig-sample}
\end{center}
\end{figure}

Our method of choice is the  previously used approach \cite{Siusys:2014_NL,Sadowski:2017_Nanoscale,Kaleta:2019_NL}, adopted from ref.~\cite{Reimer:2012_NC}.
The NWs-containing surface of the sample is covered by poly(methyl methacrylate) (PMMA, the commonly used e-beam resist) and then immersed into liquid nitrogen.
Due to a sufficiently large difference in thermal contraction the PMMA layer peels off from the semiconducting substrate taking the NWs with itself.
The remarkable advantage of this approach is that during peeling the PMMA flake reaps only the NWs, so all the planar deposits, in particular the parasitic (Ga,Mn)As-like layer, are left on the substrate.
A sufficient effectiveness of the process is exemplified in Fig.~\ref{Fig-sample}.
We acknowledge that the vast majority of the NWs has been transferred.
The NWs constitute the black substance seen through the otherwise transparent PMMA as indicated in part (b) of Fig.~\ref{Fig-sample}.

The specimens fabricated this way are very robust, withstand multiple cooling down to single-Kelvin temperatures without any noticeable degradation of the magnetic response.
Equally importantly, the rigidity of PMMA assures also the original orientation of the NWs, so this method of transferring renders also the measurements of the magnetic anisotropy possible.

Finally, we remark that PMMA is a weak diamagnet with a small admixture of a weak paramagnetism at low $T$.
The whole $T$- and $H$-dependent contribution of PMMA has been evaluated beforehand and adequately subtracted from the results presented in further parts of the report.

\subsection{ Magnetic measurements}

Magnetization measurements are carried out in a MPMS XL Superconducting Quantum Interference Device (SQUID) magnetometer equipped with a low field option.
A customary designed degaussing procedure followed by the soft quench (the "magnet reset" option of the MPMS magnetometer) of the superconducting magnet was executed prior the low field ($H \leq 1$~kOe) measurements to assure smaller than 0.2~Oe magnitude of $H$ in the sample chamber when desired.
Actually, the soft quench of the SQUID's superconducting magnet is routinely performed prior to all zero-field measurements, such as the thermo-remnant magnetization (TRM), the measurement of the decaying remnant moment on increasing $T$.
Long Si strips facilitate the samples support in the magnetometer chamber and we strictly follow the experimental code and data reduction detailed in Ref.~\cite{Sawicki:2011_SST}.

\section{Results and discussion}

\begin{figure}[t!]
\begin{center}
\includegraphics[width = 14.0cm]{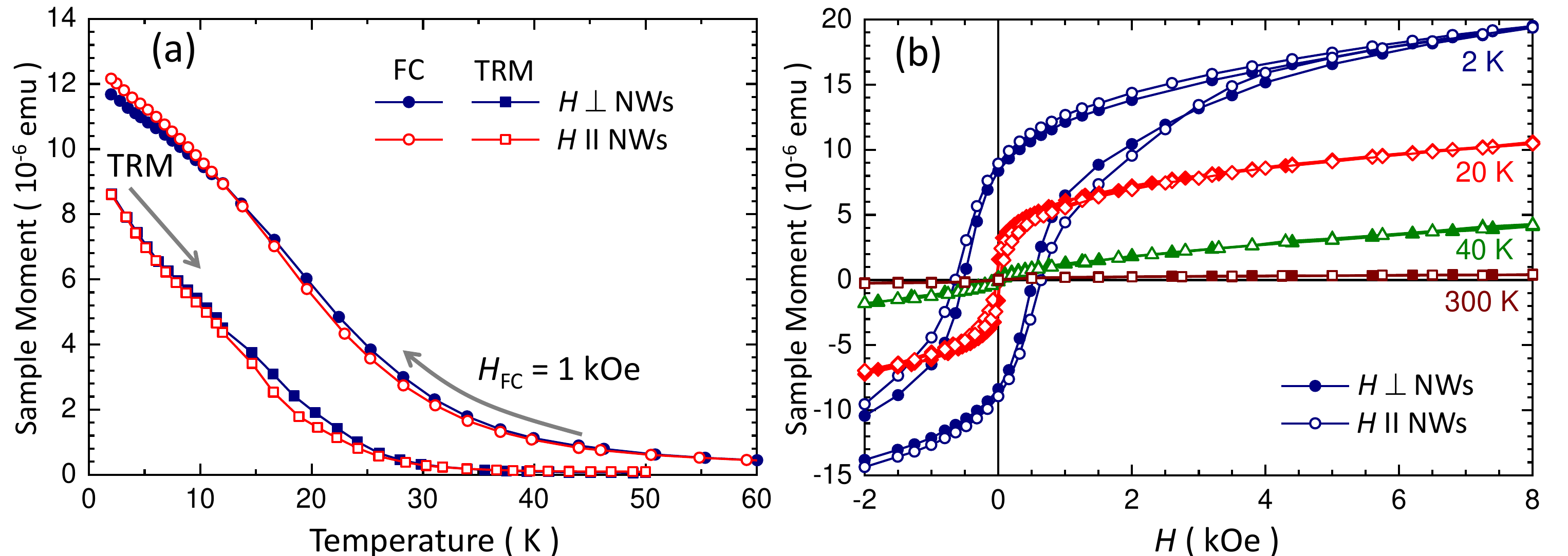}
\caption{(Color on line) (a) Temperature dependence of the magnetic moment of the wz-(Ga,Mn)As nanowire sample measured during cooling at magnetic field $H_{\mathrm{FC}} = 1$~kOe and at remanence (TRM). (b) Magnetic hystereses at selected temperatures of the same sample. All measurements are performed with two orientations of the nanowires with respect to $H$, as indicated in the labels.}
\label{Fig-MOverView}
\end{center}
\end{figure}

We start the survey of the magnetic results from the basic magnetic characterization of the NWs.
Figure~\ref{Fig-MOverView} collects low-$T$ (panel a) and low field (panel b) results, which in the case of thin layers of the parent zb-(Ga,Mn)As \cite{Sawicki:2006_JMMM} as well as in the case of NWs \cite{Siusys:2014_NL,Sadowski:2017_Nanoscale}  proved very informative about the material. 
The presence of a surge of the magnetic moment $m$ seen in panel (a) below some 40~K  during field cooling of the sample at $H = 1$~kOe, the accompanied strong non-linear response around $H=0$ during field dependent studies, and the presence of a sizable remnant moment and open hysteresis $m(H)$, panel (b),  instruct us that a strong FM spin-spin coupling develops at low temperatures in the NWs.
The strength of this coupling is particularly well seen through a large magnitude of the remnant moment below 5~K, which is about 3/4 of the field cooled one, Fig.~\ref{Fig-MOverView}~(a).

On the other hand, the features seen in Fig.~\ref{Fig-MOverView}~(a) are considerably more smeared than those observed in zb-(Ga,Mn)As exhibiting a Curie temperature $T_{\mathrm{C}}$ of about 30~K (cf. Fig.~3 in ref.~\cite{Gluba:2018_PRB} or Fig.~1 in ref.~\cite{Medjanik:2021_PRB}).
Moreover, the NW sample does not exert any worthwhile magnetic anisotropy.
Given the strong one dimensional character of the individual NW [the wz-(Ga,Mn)As nano-tubes] and their perfect spatial alignment provided by the  immobilization in solid PMMA, one should expect a strong uniaxial shape magnetic anisotropy, providing the FM coupling in the wz-(Ga,Mn)As assumed a homogeneous and long range form.

\begin{figure}[t!]
\begin{center}
\includegraphics[width = 8.0cm]{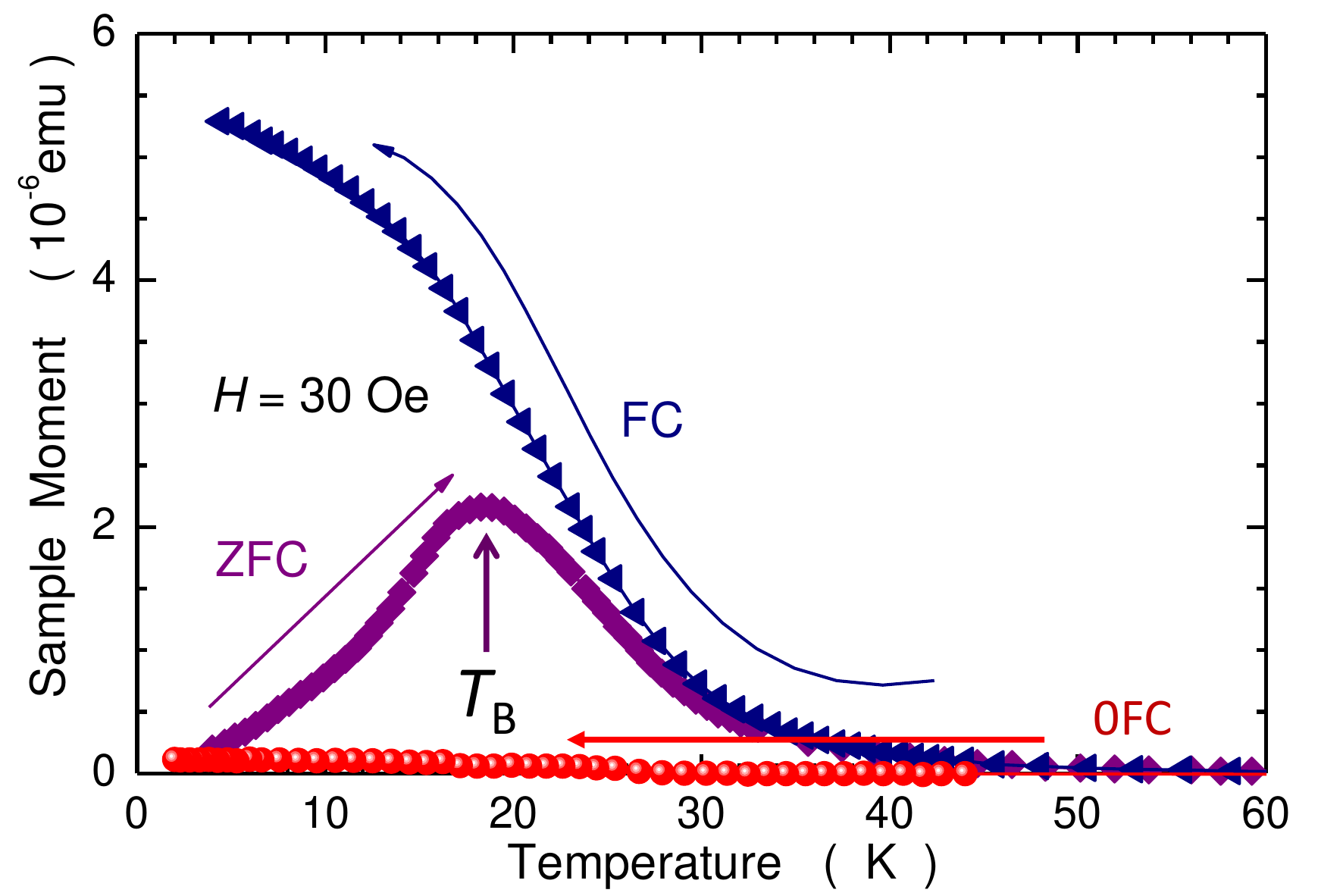}
\caption{(Color on line) Zero field cooled (ZFC, diamonds) and field cooled (FC, triangles) magnetization of the wz-(Ga,Mn)As measured at magnetic field $H = 30$~Oe. Bullets mark the temperature dependence of the magnetic signal of sample during cooling at $H=0$. The thick arrow marks the position of the blocking temperature $T_{\mathrm{B}}$.}
\label{Fig-SP}
\end{center}
\end{figure}

So, a relevant question arises whether the homogenous ferromagnetic phase forms in wz-(Ga,Mn)As?
To start unraveling this issue an augmented version of the time honored zero filed cooled (ZFC) and field cooled (FC)  measurements are performed.
This time, in a search of the spontaneous magnetization, we record also the $T$-dependence of the magnetic signal of the NWs sample during cooling at $H=0$.
We dub this measurement as 0FC.
Results of the all three measurements are shown in Fig.~\ref{Fig-SP}.


Starting from the ZFC-FC pair, we recognize the classical picture corresponding to the blocked superparamagnetic (SP) behavior.
A clear bifurcation between ZFC and FC is observed at somewhat higher temperature $T \simeq 30$~K than the blocking temperature, $T_B \simeq 18$~K, indicated as a clear maximum on the ZFC.
These SP-like properties observed in a structurally homogeneous material indicate that the wz-(Ga,Mn)As investigated here, despite quite sizable $x \simeq 6~$\%, assumes  a magnetically granular form.
That is, instead of a uniform "wall-to-wall" ferromagnetic material as it would be expected to see in zb-(Ga,Mn)As, the FM order in these wz shells is spatially limited to some mesoscopic volumes.
Mn atoms coupled in these parts of the material exert collectively a large magnetic moment, "macrospins", which develop only below a certain  temperature.
It will be evaluated  later.

The magnetically granular form of the wz-(Ga,Mn)As is substantiated further by the lack of any spontaneous moment exerted by the sample  during the initial 0FC cooling.
This is in a stark contrast to planar zb-(Ga,Mn)As, which upon 0FC exhibits a rapid surge of magnetic signal below its Curie temperature, $T_{\mathrm{C}}$ (cf. eg.~Fig.~3 in ref.~\cite{Sawicki:2010_NP} or Fig.~4 in ref.\cite{Gluba:2018_PRB}).
The magnitudes of those spontaneously formed magnetizations are very close to the saturation ones for each of these samples, indicating that the spontaneous magnetizations are formed upon passing $T_{\mathrm{C}}$ at $H = 0$.
Nothing like this is seen in Fig.~\ref{Fig-SP}, despite the fact that the Mn content in the (Ga,Mn)As nano-tubes exceeds about 4 times that of the sample from ref.~\cite{Gluba:2018_PRB}.
We understand that in the absence of external field the macrospins  point at random orientations, 
and so, as an ensemble, they exert no net magnetic moment.

\begin{figure}[t!]
\begin{center}
\includegraphics[width = 8.5cm]{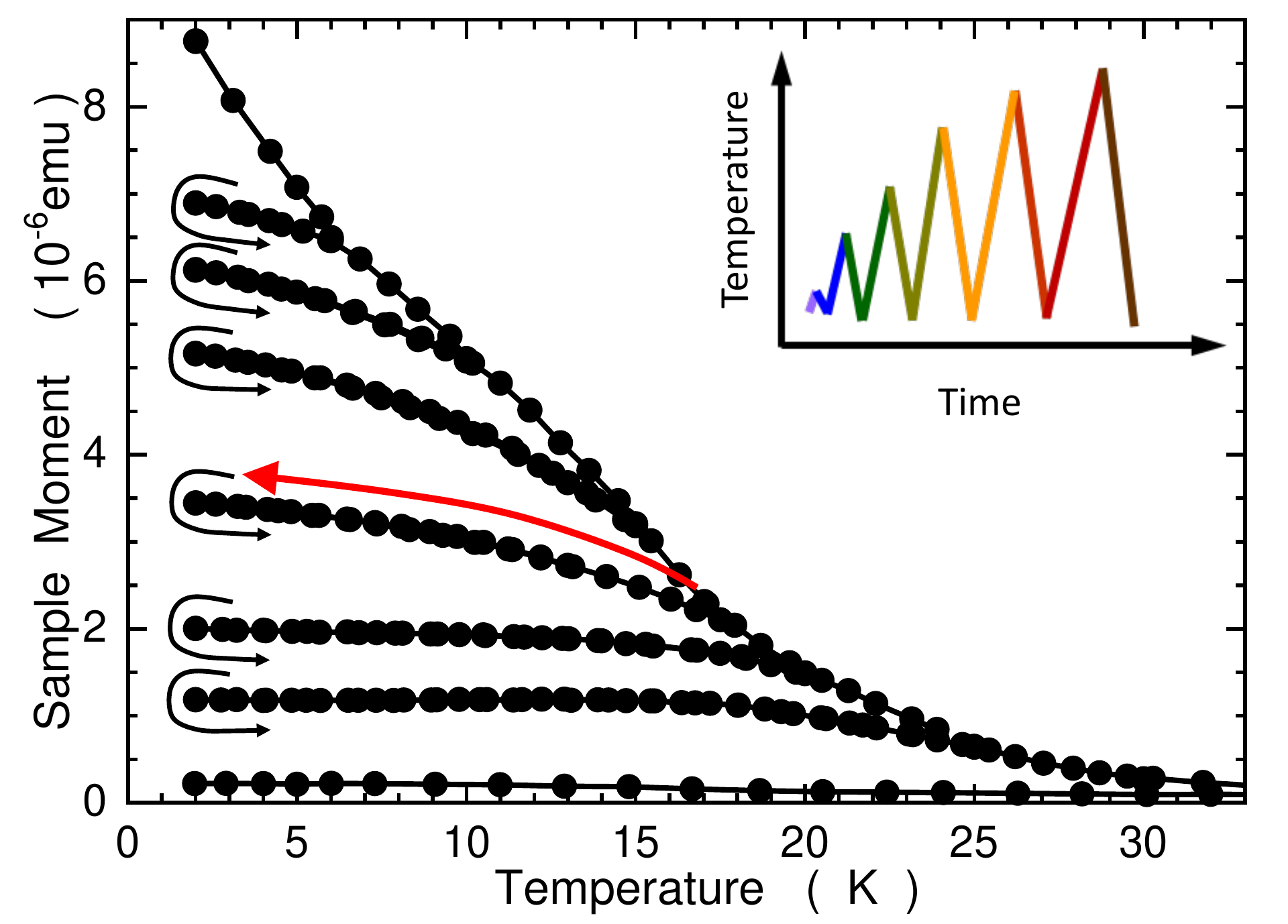}
\caption{Thermal cycling of thermo-remnant magnetization (mTRM) during which the sample undergoes a process of repeated  warmings and coolings at zero magnetic field, as schematically sketched in the inset. The red thick arrow exemplifies the fact that the concave curvature of any branch of the mTRM is perfectly reproduced when the sample is re-cooled.}
\label{Fig-mTRM}
\end{center}
\end{figure}

The smoking gun experimental evidence confirming the \emph{magnetically} inhomogeneous structure of wz-(Ga,Mn)As is depicted in Fig.~\ref{Fig-mTRM}.
This variant of standard TRM, mTRM, allows to distinguish between a long range FM order and the granular form of the magnetic constitution of the sample \cite{Mamiya:2007_JMMM}, and even to quantify the relative contributions of such two components - if they are simultaneously present in the sample \cite{Sawicki:2010_NP}.
The mTRM measurement, like the TRM one, starts from bringing the sample to the remanence at the base temperature, but instead of continuous warming, the sample is warmed up in discrete stages, according to the pattern sketched in the inset to Fig.~\ref{Fig-mTRM}.
In mTRM the sample undergoes a process of repeated  warmings and coolings, achieving a progressively larger $T$ at each consecutive stage of the process.
The  mTRM pattern presented in Fig.~\ref{Fig-mTRM} does resemble a great deal the ladder-like structure expected  for a fine FM particles system, cf.~Fig.~2~(b) of ref.~\cite{Mamiya:2007_JMMM}.
And so it decisively confirms the  magnetically granular form of magnetism in the wz-(Ga,Mn)As shells.


There is, however, one important difference between these two results.
In ref.~\cite{Mamiya:2007_JMMM} the steps of the mTRM "ladder" are flat and they run horizontally.
Here, a ceratin downward slope is seen at low temperatures and, most importantly, a sizable curvature develops at higher temperatures.
We assign these discrepancies to the profound differences of the material systems studied here and in ref.~\cite{Mamiya:2007_JMMM}.
In the latter case a nanocrystalline form of Fe$_3$N is investigated at $T < 50$~K~$ << T_{\mathrm{C}}^{\mathrm{Fe}_3\mathrm{N}} \simeq 650$~K  \cite{Leineweber:1999_JALCOM,Navarro-Quezada:2020_Materials}.
Therefore the magnitude of Fe$_3$N macrospins  can be regarded independent of $T$ below 50~K and the effects observed there are due to (size dependent) temperature-agitated relaxation of macrospins to the equilibrium state.
At the equilibrium the macrospins get randomly oriented and they remain this way upon subsequent cooling at $H = 0$.
However, when the  experimental range of $T$ extends up to $T_{\mathrm{C}}$, like in Fig.~2~(a) of ref.~\cite{Mamiya:2007_JMMM}, that is when mTRM of a low-$T_{\mathrm{C}}$ \emph{bulk} FM sample is measured, a well developed concave curvature appears on warming.
In this case the remanent magnetization follows the $T$-dependence of the spontaneous magnetization ($M_{\mathrm{sp}}$).


Such a concave curvature of mTRM is marked by the curved red arrow in  Fig.~\ref{Fig-mTRM}, indicating that in our experiment we probe both regions of $T$ simultaneously.
The very much ladder-like structure reconfirms us about the magnetic granular form, whereas the pronounced concave curvature, exhibited in particular on \emph{re-cooling},  underpins the leading contribution of the $T$-dependence of the magnitude of the macrospins' moments. 
In other words, such a strong curvature of mTRM signalizes a close proximity to the internal $T_{\mathrm{C}}$ of the macrospins.


At this point an important question arises why the long range FM order has been replaced by SP macrospins in the otherwise structurally homogeneous wz-(Ga,Mn)As shells with $x = 6$\%.
We start from remarking that the SP-like contribution to $M$ in both zb- and wz-(Ga,Mn)As was noted previously \cite{Sawicki:2010_NP,Siusys:2014_NL,Chen:2015_PRL,Yuan:2017_PRM,Gluba:2018_PRB,Yuan:2018_JPCM}, in fact the awareness of this fact is constantly growing \cite{Sawicki:2018_PRB}. 
Since the material contains a rather sufficient concentration of Mn to support the global FM ordering, it has to be a too low  density of  mobile holes to spread the coupling uniformly across the whole volume of the (Ga,Mn)As shells.
Interestingly, a low hole density does not completely preclude the formation FM in the mean field manner.
At the vicinity of the metal-insulator-transition (MIT) the FM coupling may be mediated  by weakly localized holes (WLH)  \cite{Dietl:2000_S,Ferrand:2001_PRB,Kepa:2003_PRL}.
They operate only on small (mesoscopic) volumes which size and distribution is set by large spatial fluctuation of the local density of states (LDOS) \cite{Dagotto:2001_PR,Richardella:2010_S}.
These WLH can mediate the FM coupling, but only within their localization length, which remains non-zero even when the average hole density is smaller than the critical one.
In other words, the FM order forms in these hole-enriched volumes (FM bubbles) and they are the source of the local magnetization $\rightarrow$ the SP macrospins.

So, this is the existence of such irregular and isolated puddles in which WLH reside \cite{Richardella:2010_S}, which gives rise to FM spin-spin coupling and so the formation of SP macrospins.
Such a situation is met for low and very low $x$ \cite{Yuan:2017_PRM,Gluba:2018_PRB} as well as in high-$x$ (Ga,Mn)As which got deprived of itinerant holes either by a strong electrical compensation or by interface depletion effects \cite{Sawicki:2010_NP,Wurstbauer:2008_APL,Siusys:2014_NL,Chen:2015_PRL,Sawicki:2018_PRB}.

Therefore the next relevant question  is why the hole concentration in our wz-(Ga,Mn)As shells got so low at the first place.
We start from a notion that the growth of only 10--20~nm thin (Ga,Mn)As generally promotes an increase of $T_{\mathrm{C}}$ \cite{Proselkov:2012_APL}.
But here the effect is the opposite.
There is no global coupling at all.
We argue that it is an excess of As supply during the NWs growth which is responsible for so low hole density in our (Ga,Mn)As shells.

The geometry of NWs growth in the MBE chamber is such that the arriving atoms from the effusion cells meet the NWs walls at high incident angles, only  around 20\degree\, to the NWs surface. 
This reduces the effective flux to about $\sin(20\degree) = 34$\% of its calibrated values for the planar growth.
This number has to be further reduced about twice due to the "shadowing effect": only one side of the NWs is exposed to the impinging fluxes at a given time.
However, these considerations apply mostly to Ga and Mn fluxes, since these atoms are incorporated into the forming crystals at locations which they hit.
The same applies to the planar (In,Ga)As base formed during the growth of  the wz-(In,Ga)As cores.
This is the mechanism which gave rise to the formation of the parasitic (Ga,Mn)As-like layer in-between the NWs that we needed to get rid during the preparation of the specimen(s).
On the other hand, arsenic atoms are not well adsorbed on this surface,  they rather bounce off back towards the NWs where they finally get adsorbed in quantity a way greater than desired.
As the result, the NWs are grown in an effective excess of As what leads to the formation of numerous As antisites.
Since they are double donors in III-V lattices the concentration of holes gets dramatically reduced.
The massive scale of this effect was documented in \cite{Wurstbauer:2008_APL}, where the supply of surplus As during the growth of (110) (Ga,Mn)As layer with $x = 6$\% resulted in the complete quench of the long range order.


\begin{figure}[t!]
\begin{center}
\includegraphics[width = 14cm]{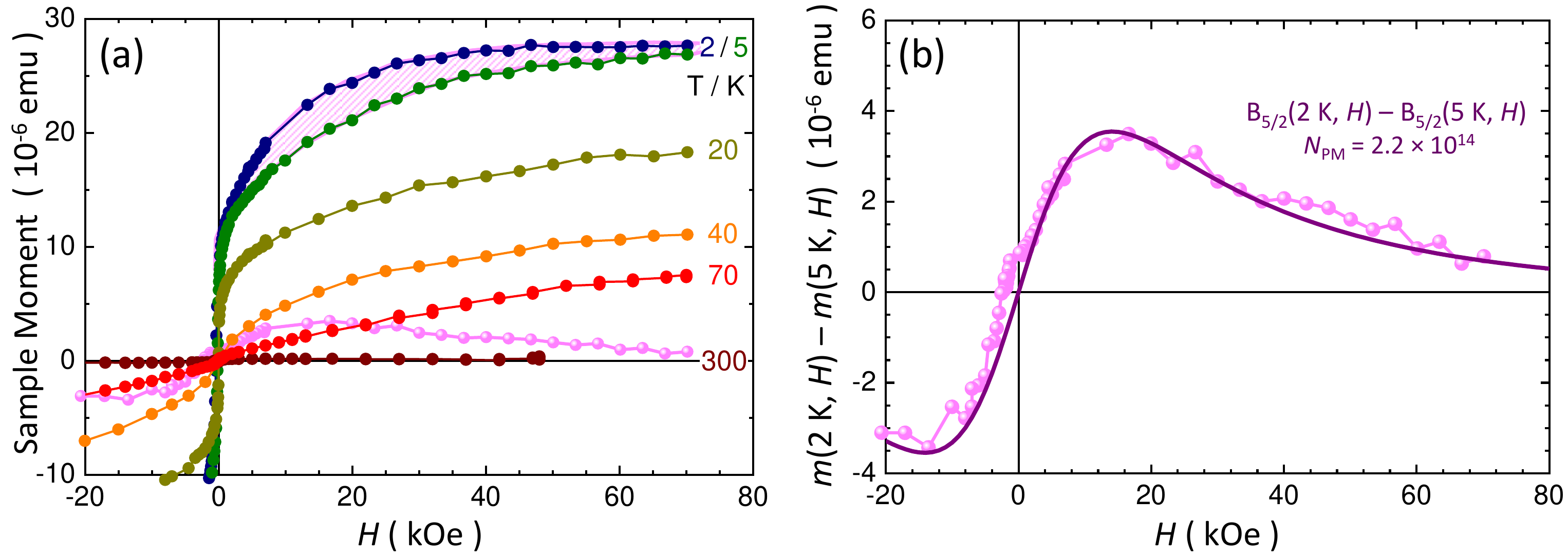}
\caption{(Color on line) (a) Magnetic isotherms $m(H)$ at some selected temperatures (indicated in the panel). The hatched area and the light hue bullets mark the difference between $m(H)$ at 2 and 5~K. (b) The same difference between $m(H)$ at 2 and 5~K re-plotted in an expanded scale (bullets) with an equivalent difference between Brillouin functions $B(H)$ calculated for $J=S=5/2$ (solid line). The matching factor between the theory and experiment, $N_{\mathrm{PM}}$, yields the number of paramagnetic spins in the sample. }
\label{Fig-PM}
\end{center}
\end{figure}


The presence of strong fluctuations of LDOS spells not only the existence of mesoscopic volumes enriched with high density of holes.
Simultaneously,  the rest of the material should be nearly deprived of holes, so clearly visualized in \cite{Richardella:2010_S}.
In these regions the Mn spins stay uncoupled, so they should exert a paramagnetic (PM) response. 
We quantify this PM component by employing a concept developed originally for GaN:Fe composite system \cite{Pacuski:2008_PRL,Navarro:2010_PRB} and proven effective in other transition metal doped DMS systems \cite{Gas:2021_JMMM_Cr}.

In order to quantify the PM contribution we measure magnetic isotherms $m(H)$ at a possibly large magnetic field range and evaluate  the differences between two low-$T$ ones.
These temperatures should be much smaller than the spin-spin coupling temperature of the non-PM part of the system, so the latter magnetization could be regarded as $T$-independent at very low temperatures.
The collected magnetic isotherms are shown in Fig.~\ref{Fig-PM}~(a).
For the evaluation we take $m(H, 2$~K) and $m(H, 5$~K).
The difference of interest is hatched in the same Fig.~\ref{Fig-PM}~(a) and re-plotted in part (b) by  bullets of light hue.
Next, this difference is approximated by  the difference of two Brillouin functions $g \mu_{\mathrm{B}} N_{\mathrm{PM}}J \times \Delta B_J(T,H)$.
Here $B_J(T,H)$ assumes the textbook form of the Brilloiun function and the values of $T$ in the arguments of $B_J$ are the same as in experiment.
We take $J = S=5/2$ and the corresponding Mn Land\'e factor $g=2.0$ \cite{Matsukura:1998_PRB,Krstajic:2004_PRB}.
In this approach the number of the PM spins in the specimens,  $N_{\mathrm{PM}}$, is the only adjustable parameter.
The dark solid line, passing so convincingly on top of the experimental scatter of points in Fig.~\ref{Fig-PM}~(b), is obtained for $N_{\mathrm{PM}} = 2.2\times 10^{14}$.

\begin{figure}[t!]
\begin{center}
\includegraphics[width = 14cm]{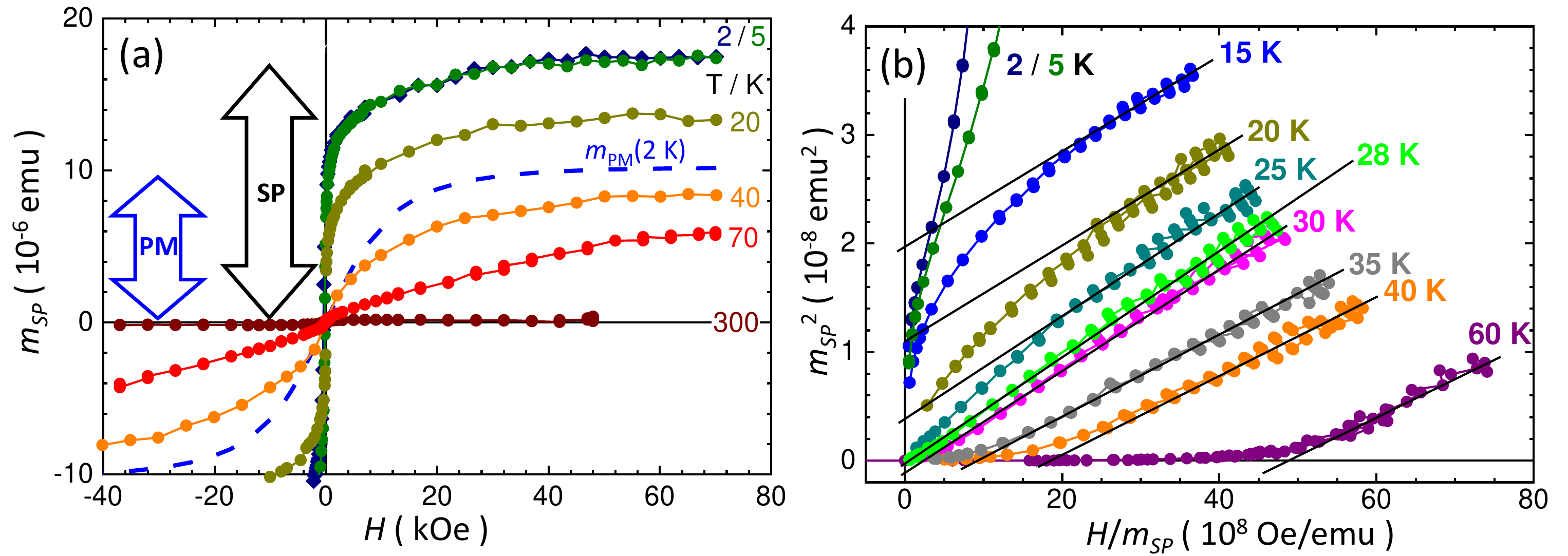}
\caption{(Color on line) (a) Magnetic field, $H$, dependence of the superparamagnetic (SP) contribution, $m_{\mathrm{SM}}$, to the total magnetic response, $m$, of the wz-(Ga,Mn)As at selected temperatures (bullets). The dashed line marks the maximum paramagnetic (PM) contribution of Mn$^{2+}$ ions, that is calculated at 2~K. The two thick arrows indicate the magnitudes of PM and SP components to $m$. (b) The Arrott plot for the SP part of $m$, $m_{\mathrm{SP}}$.   }
\label{Fig-Separa-Arrott}
\end{center}
\end{figure}

Having established the magnitude of $N_{\mathrm{PM}}$ we can calculate the magnitudes of the PM contribution to $m$, $m_{\mathrm{PM}}(T,H) = g \mu_{\mathrm{B}} N_{\mathrm{PM}} (5/2) B_{5/2}(T,H)$ at any given $T$ and $H$ and subtract it from the original data to obtain the sole contribution brought by the locally mediated coupling. 
These dependencies are presented in Fig.~\ref{Fig-Separa-Arrott}~(a), where magnetic hystereses specific to the SP-like component $m_{\mathrm{SP}}(H)$ are plotted for some selected temperatures.
The lack of visible differences between $m_{\mathrm{SP}}(H, 2$~K) and $m_{\mathrm{SP}}(H, 5$~K) confirms the correctness of the PM-separation procedure.
For comparison, the magnitude of  $m_{\mathrm{PM}}(H)$ at $T=2$~K is added (the dashed line).
This yields the ratio of these two components to $m$  to be as 63 to 37, or nearly 2:1 in favor for that part of the material which is coupled by the WLH.
Relying on the assumption of the uniform Mn distribution within the (Ga,Mn)As nano-tubes this results yields the same, about 2/3, FM volume fraction for this magnetically phase separated system.

After separating out the PM component to $m$ we are in the position  to assess the magnitude of the intra-$T_{\mathrm{C}}$ of the macrospins.
To this end we employ yet another time honored method, the Arrott-plot concept \cite{Arrott:1957_PR}, regardless the fact that in the high quality metallic and single magnetic phase zb-(Ga,Mn)As a more involved approach, the Kouvel-Fisher plots \cite{Kouvel:1964_PR},  would be far more appropriate \cite{WangMu:2014_APL}.
On the other hand, on the account of the accumulated evidences we know that the macrospins are not characterized by a unique and well defined intra-$T_{\mathrm{C}}$.
On the contrary, the vague nature of WLH and the strongly fluctuating character of LDOS suggest a rather wide spectrum of these magnitudes, so our attempt has only a semi-quantitative value.

Figure \ref{Fig-Separa-Arrott}~(b) collects the plots of $m^2$ vs.~$H/m$ for some selected temperatures, with an emphasis put at the vicinity of 30~K.
The linearly extrapolated experimental data point to about 28~K as the statistically most significant magnitude  of the  intra-$T_{\mathrm{C}}$ of the SP part of the sample.
For certainly, there has to exist a spread of these intra-$T_{\mathrm{C}}$ magnitudes, at least the generally smeared shape of TRM suggests so.
It remains to be seen whether such a shape of TRM as seen in Fig.~\ref{Fig-mTRM} is the hallmark of the dominance of the independent macrospins in the magnetic constitution of (Ga,Mn)As.
Or whether the wz structure has anything to do with it.
Most of the (Ga,Mn)As samples were grown to maximize their "long range" $T_{\mathrm{C}}$, so in this respect the wz-(Ga,Mn)As shells  investigated here provide a very important play-ground to test the very nature of the spin-spin coupling in this still very important material class.
According to the authors' knowledge this is the first experimental assessment of the intra-$T_{\mathrm{C}}$ of the macrospins forming in (Ga,Mn)As with a substantial deficiency of holes.

\section{Conclusion}

In conclusion, the investigated high crystalline quality wurtzite (Ga,Mn)As has been obtained in a form of quasi-1D shells (nanotubes) overgrown by heteroepitaxy on wurtzite (Ga,In)As nanowires using molecular beam epitaxy.
A carefully prepared specimen exhibited a strong magnetic response below about 30~K deceivingly resembling a good FM material, namely a non-zero remanence and magnetic hystereses were present.
Zero field cooled and field cooled measurements indicated a granular form of the magnetism, which has been decisively confirmed by the thermal cycling of the thermo-remnant magnetization and by a complete absence of the spontaneous magnetization.
On this account the ferromagnetic-like features have been assigned to the dynamical slow-down characteristic to blocked superparamagnetic volumes - macrospins, formed in these parts of the material which are populated by weakly localized holes.
Using high field results it has been evaluated that these volumes constitute about 2/3 of the volume of the (Ga,Mn)As whereas in the remaining 1/3 only paramagnetic Mn ions reside.
By establishing the number of the uncoupled ions the two contributions were separated and the Arrott plot method applied to the superparamagnetic part yielded the first experimental assessment of the magnitude  of the spin-spin coupling temperature within the macrospins in (Ga,Mn)As,  $T_{\mathrm{C}}=28$~K.
The existence of such a patchwork of spontaneously magnetized and nonmagnetic regions (volumes) in the material has been assigned to  a relatively low hole concentration caused by numerous arsenic antisites formed due to a specific geometry of the growth of the shells on the nanowire template.

\section*{Acknowledgments}
The authors are thankful to A. Siusys for the preparation of the specimens. J.S. acknowledges partial support by Carl Tryggers Stiftelse (Sweden) under the project CTS 16:393.


\end{document}